\documentclass[sigconf]{acmart}
\AtBeginDocument{%
  }

\copyrightyear{2026}
\acmYear{2026}
\setcopyright{cc}
\setcctype{by}
\acmConference[SIGIR '26]{Proceedings of the 49th International ACM SIGIR Conference on Research and Development in Information Retrieval}{July 20--24, 2026}{Melbourne, VIC, Australia}
\acmBooktitle{Proceedings of the 49th International ACM SIGIR Conference on Research and Development in Information Retrieval (SIGIR '26), July 20--24, 2026, Melbourne, VIC, Australia}
\acmDOI{10.xxx}
\acmISBN{xxx}
\usepackage{balance}
\begin{document}

%% allowing the author to define a "short title" to be used in page headers.\title{Bridging Passive Selection and Active Expression: A Heterogeneous Behavior Modeling Framework for Conversation Starter Recommendation }

\title{Bridging Passive and Active: Enhancing Conversation Starter Recommendation via Active Expression Modeling}

\author{Yiqing Wu}
\authornote{equal contribution, $\dagger$ corresponding author}


\affiliation{%
  \institution{Bytedance}
  \city{Beijing}
  \country{China}
}
\email{iwu_yiqing@163.com}

\author{Haoming Li}
\authornotemark[1]

\affiliation{%
  \institution{Bytedance}
  \city{Beijing}
  \country{China}}
\email{lihaoming.cs@bytedance.com}

\author{Guanyu Jiang}
\affiliation{%
  \institution{Bytedance}
  \city{Beijing}
  \country{China}}
\email{jiangguanyu.jgy@bytedance.com}

\author{Jiahao Liang}
\affiliation{%
  \institution{Bytedance}
  \city{Beijing}
  \country{China}
}
\email{liangjiahao.24@bytedance.com}

\author{Yongchun Zhu}
% \authornote{corresponding author}
\authornotemark[2]
\affiliation{%
  \institution{Bytedance}
  \city{Beijing}
  \country{China}}
\email{zhuyc0204@gmail.com}

\author{Jingwu Chen}
\authornotemark[2]
\affiliation{%
  \institution{Bytedance}
  \city{Beijing}
  \country{China}}
\email{chenjingwu@bytedance.com}

\author{Feng Zhang}
\affiliation{%
  \institution{Bytedance}
  \city{Beijing}
  \country{China}}
\email{feng.zhang@bytedance.com}

\renewcommand{\shortauthors}{Yiqing Wu et al.}
\newcommand{\modelname}{PA-Bridge}
\begin{abstract}

% Large Language Model (LLM)-driven conversational search is shifting retrieval toward proactive dialogues, where Conversation Starters are widely deployed to recommend personalized opening queries.

Large Language Model (LLM)-driven conversational search is shifting information retrieval from reactive keyword matching to proactive, open-ended dialogues. In this context, Conversation Starters are widely deployed to provide personalized query recommendations that help users initiate dialogues.
Conventionally, recommending these starters relies on a closed "exposure-click" loop. Yet, this feedback loop mechanism traps the system in an echo chamber where, compounded by data sparsity, it fails to capture the dynamic nature of conversational search intents shaped by the open world.
As a result, the system  skews towards popular but generic suggestions.
In this work, we uncover an untapped paradigm shift to shatter this harmful feedback loop: harnessing user "free will" through active user expressions. Unlike traditional recommendations, conversational search empowers users to bypass menus entirely through manually typed queries. The open-world intents in active queries hold the key to breaking this loop. However, incorporating them is non-trivial: (1) there exists an inherent distribution shift between active queries and  formulated starters. (2) Furthermore, the "non-ID-able" nature of open text renders traditional item-based popularity statistics ineffective for large-scale industrial streaming training.
To this end, we propose \textbf{Passive-Active Bridge (PA-Bridge)}, a novel framework that employs an adversarial distribution aligner to bridge the distributional gap between passively recommended starters and active expressions. Moreover, we introduce a semantic discretizer to enable the deployment of popularity debiasing algorithms. Online A/B tests on our platform, demonstrate that PA-Bridge significantly boosts the Feature Penetration Rate by 0.54\% and User Active Days by 0.04\%.

\end{abstract}


% \ccsdesc[500]{}
% \ccsdesc[300]{}
% \ccsdesc{}
% \ccsdesc[100]{}



\maketitle

\begin{figure}[!t]
    \setlength{\abovecaptionskip}{0pt} 
    \setlength{\belowcaptionskip}{0pt}
\centering
\includegraphics[width=0.99\columnwidth]{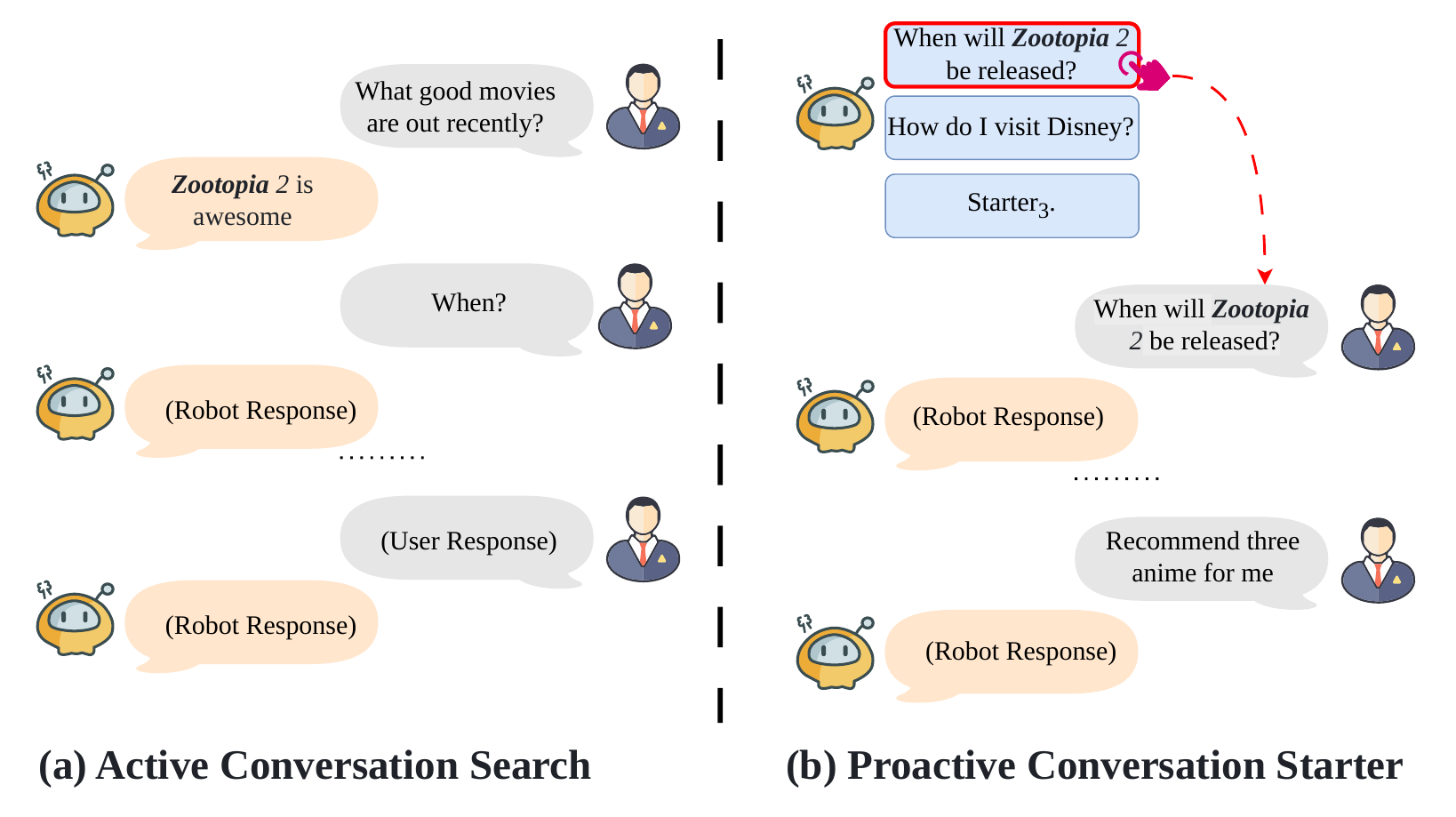}
\caption{Illustration of  Conversation Starter. It proactively provides queries  to guide users to start a conversation }
\label{fig:example}

\end{figure}

\section{Introduction}

Large Language Model (LLM)-based conversational search is superseding traditional search boxes, and shifting the interaction paradigm from reactive retrieval to proactive engagement. In this context, Conversation Starters are widely deployed to  provide personalized, proactive queries that help users initiate dialogues, as shown in Fig. ~\ref{fig:example}. Conventional industry practices model this task through a classic recommendation paradigm: retrieving candidate starters from a pre-constructed corpus and ranking them based on historical engagement  \cite{ahmad2018multi,chen2020incorporating,lee2024enhanced}. In practice, this forms an ``exposure-click-training" cycle, where the model relies exclusively on user feedback from its own displayed items.

However, this conventional self-looping mechanism harbors a systemic flaw. 
% The system learns user preferences solely from interaction logs generated by its own previous exposures.  
The "exposure-click-training" cycle implies that what users click is heavily dictated by what the system chooses to push, rather than what they genuinely desire. Furthermore, this structural flaw becomes particularly fatal in the context of conversation starter. Unlike traditional static item recommendations, a user's conversational intent is inherently open-ended and dynamically driven from the open world. Consequently, when encountering such open-ended query intents, the closed self-loop system  fails to perceive and capture the shifting user query needs. As a result, we observe the model tends to fall back on suggesting common, trivial starters at the expense of true personalization. Actually, the top 1\% of popular starters account for 70\% of the total exposure in our system.
% How to solve this problem still is an open problem.

To shatter this harmful feedback loop, a powerful  `hammer' is urgently needed. In fact, a critical signal has been chronically overlooked: the user's active query. We realize that  \textit{implicit
click signals merely scratch the surface of user interest, whereas active queries explore open-ended intent frontiers.}  
While traditional platforms like Amazon or Netflix operate like enclosed Hakoniwa (box garden) games—where user interests are learned within a strictly bounded set of predefined items—modern conversational search has evolved into a highly autonomous Open World. In this open world, users no longer just react to what is pushed; they actively explore open-ended intent frontiers through free-text queries. However, conventional  models remain trapped in the   `Hakoniwa' mindset, focusing solely on sparse reactive signals and ignoring the rich, explicit interests expressed in the broader open-world environment.

However, harnessing these open-world queries for starter recommendation within a closed-world corpus is non-trivial.  The bridge  between them is broken by distribution shifts and their non-ID-able nature. Specifically, manually typed queries are open-world and unformulated, whereas starters are selected from a closed, formulated corpus. Furthermore, incorporating active queries introduces  a severe popularity bias, as user queries naturally follow a long-tail distribution. Eliminating this bias is crucial for maintaining system health. However, achieving this in industrial streaming training poses a significant challenge. Standard debiasing methods fundamentally rely on discrete item-ID popularity statistics \cite{yi2019sampling}. Consequently, they cannot be directly extended to handle the unstructured, highly diverse, and intrinsically "non-ID-able" space of active user queries.

To bridge the gap between passively  suggested starters and actively expressed queries, we propose a novel framework, \textbf{Passive-Active Bridge (PA-Bridge)}. PA-Bridge posits that despite surface differences, active queries and suggestions share a common underlying intent. By projecting them into a unified semantic space, we can effectively align and leverage these distinct signals. PA-Bridge comprises two core components: Distribution Aligner: To construct this unified semantic space, we introduce an Adversarial Learning mechanism. A discriminator is employed to distinguish the query source, forcing the encoder to strip away noise and learn domain-invariant representations; Semantic Discretizer: While a unified space theoretically allows for density sampling, directly modeling high-dimensional continuous distributions is computationally intractable. We employ RQ-VAE to quantize the continuous space into discrete semantic clusters.
This transforms infinite active queries into finite 'Semantic IDs', enabling us to approximate  density estimation through efficient discrete probability sampling, thereby facilitating effective debiasing.

PA-Bridge has been successfully deployed on our platform. Experimental results show that our approach effectively leverages active queries, yielding a $6.27\%$ offline gain and a $0.04\%$ online lift in User Active Days. Our contributions are as follows:
\begin{itemize}
    \item We pioneer the integration of active user expressions into  proactive query recommendation to alleviate the  harmful feedback loop and the echo chamber effect.
    \item We propose PA-Bridge, which employs adversarial alignment and semantic discretization to bridge distribution gaps and enable ID-free debiasing in a unified latent space.
    \item We validate the effectiveness and robustness of our method through extensive offline and online evaluations in a real-world production environment.
\end{itemize}

\section{Methodology}
\begin{figure}[!hbtp]
    \setlength{\abovecaptionskip}{0pt} 
    \setlength{\belowcaptionskip}{0pt}
\centering
\includegraphics[width=0.99\columnwidth]{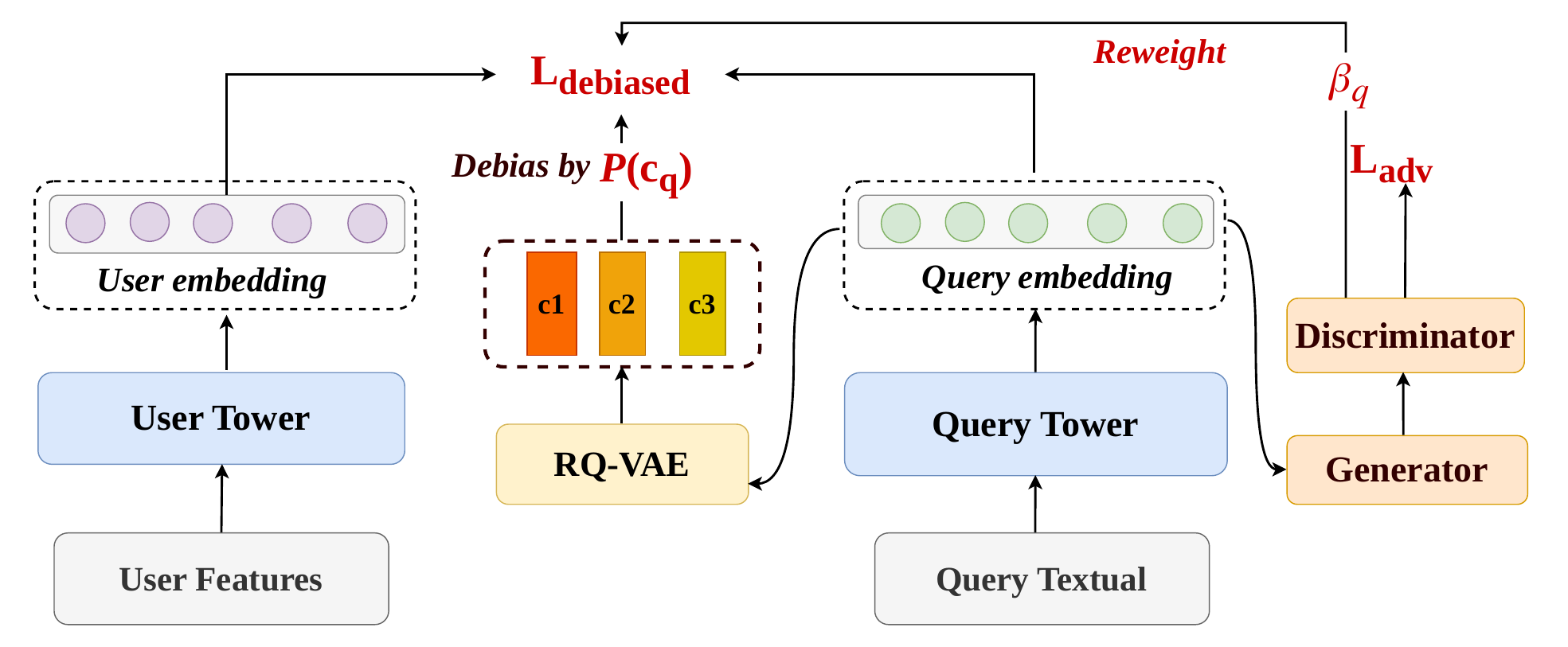}
\caption{Overall architecture of our PA-Bridge.}
\label{fig:framework}

\end{figure}

In this section, we present our \textbf{PA-Bridge}, as shown in Fig.~\ref{fig:framework}. 
First, we provide the formal \textbf{Problem Definition} in Sec.~\ref{sec:probme formulation}. 
Next, we describe the \textbf{Basic Model Architecture} in Sec.~\ref{sec:Base Model Architecture}, which serves as the backbone of our system. 
To effectively leverage active queries, we introduce two core components: 
in Sec.~\ref{sec:Distribution Aligner}, we propose the \textbf{Distribution Aligner} to bridge the semantic gap between active and  starters; 
and in Sec.~\ref{sec:Semantic Discretizer}, we present the \textbf{Semantic Discretizer} to enable unbiased learning without explicit item IDs.

\subsection{Problem Formulation}

\label{sec:probme formulation}
We formulate the ``Conversation Starter'' generation as a standard Top-$K$ recommendation task. Let $\mathcal{U}$ and $\mathcal{Q}$ denote the set of users and the corpus of pre-defined candidate starters, respectively. Given a user $u \in \mathcal{U}$, our objective is to learn a scoring function $f(u, q ; \Theta)$, parameterized by $\Theta$, to estimate the engagement probability for each candidate starter $q \in \mathcal{Q}$. Based on the computed scores, the system retrieves and returns a ranked list of $K$ starters, $S = [s_1, s_2, \dots, s_K]$ where $s_i \in \mathcal{Q}$, to the user. Furthermore, to distinguish between the data sources, we use $q^{a} \in  \mathcal{Q}^a $ and $q^{s} \in  \mathcal{Q}^s$ to denote the user's active query and the passively suggested starter.

\subsection{Base Model Architecture}
\label{sec:Base Model Architecture}
In this section, we present the basic structure of our model. To satisfy the stringent latency requirements of industrial online serving, we adopt the standard Two-Tower architecture, comprising a User Tower and a Query Tower to independently encode user and starter representations. The scoring function is defined as:
\begin{equation}
f(u, q;\Theta) = \langle \Phi_U(\mathcal{X}_u), \Phi_Q(\mathcal{X}_q) \rangle,
\end{equation}
where $\mathcal{X}_u$ and $\mathcal{X}_q$ denote the feature sets for the user and the query, respectively. $\Phi_U$ and $\Phi_Q$ represent the network of the User and Query Towers, and $\langle \cdot, \cdot \rangle$ denotes the inner product operation.

Evidently, textual features play a pivotal role in our modeling. To capture deep semantic nuances, we employ a pre-trained Language Model (LM) as a Query Tower to project all textual inputs into dense embeddings. This process is formulated as:
\begin{equation}
\label{LM_encoder}
z = \Phi_Q(\mathcal{X}_q) =\text{LM\_Encoder}(t), \quad t \in \{q^a, q^s\},
\end{equation}
where $t$ represents any arbitrary text input, including starter $q^s$, active query $q^a$.
%The resulting vector $\mathbf{v} \in \mathbb{R}^d$ serves as the foundational representation for subsequent sequence aggregation and feature interaction. 
Finally, we optimize the model using the standard cross-entropy loss with in-batch negative sampling \cite{hidasi2015session}. Given a training batch of $B$ pairs $\{(u_i, q_i^+)\}_{i=1}^B$, where $q_i^+$ is the positive query for user $u_i$, the loss is formulated as:
\begin{equation}
\label{in_batch_softmax}
\mathcal{L}_{base} = - \frac{1}{B} \sum_{i=1}^{B} \log \frac{\exp(f(u_i, q_i^+)/\tau)}{\sum_{j=1}^{B} \exp(f(u_i, q_j)/\tau)}.
\end{equation}
% where $\tau$ is a temperature hyper-parameter to scale the logits.
\subsection{Distribution Aligner}
\label{sec:Distribution Aligner}
% \subsection{Bridging the Distribution Gap}

\label{sec:Distribution Alignment}
As discussed before, the conventional "exposure-click-training" loop traps conversation starter recommendation in a closed-world echo chamber. To shatter this harmful feedback loop, we identify the user's manually typed active queries as a pivotal bridge, since they capture dynamic, open-ended intent frontiers.

A straightforward strategy to utilize active queries is to augment the training set with $q^{a}$ as auxiliary positive samples.  However, we observe a significant \textit{distribution shift} between them.  Clearly, the starters are typically well-structured, whereas users' manually typed queries tend to be unformulated and highly colloquial. Furthermore, active user queries encompass an open-world space, while our predefined starters are constrained to a closed-world one. This discrepancy inevitably leads to the well-known Sample Selection Bias (SSB) problem \cite{ma2018entire}. 

Despite this semantic gap, both active queries and passive starters fundamentally reflect the underlying user intents. Therefore, they share a common latent intent space, and projecting them into the same space can alleviate the distribution shift problem.  To achieve this, we propose an Adversarial Learning Framework. This framework primarily operates during the offline training phase and comprises three key components:
(1) \textit{Feature Extractor}: It is responsible for extracting raw textual representations from the input. We adopt the aforementioned LM Encoder (Eq.(~\ref{LM_encoder}))  as the Feature Extractor;
(2)\textit{Intent Mapper}: It serves as the Generator( $G$), designed to project these raw representations into the shared latent intent space; and
(3) \textit{Domain Discriminator} ($D$): It is employed to prevent the generator from collapsing to a trivial solution.
The discriminator is trained to classify whether the input feature representations belong to the active query $q^a$ or the recommended starter $q^s$. In this work, we simply adopt two MLPs as $D$ and $G$.
Typically, this alignment process is formulated as a classical minimax game \cite{goodfellow2014generative}.  This adversarial interplay is optimized via the following minimax objective:
\begin{equation}
\begin{aligned}
\mathcal{L}_{adv} = \min_{G} \max_{D} \mathbb{E}_{z \sim \mathcal{D}_A} [\log D(G(z))] + \mathbb{E}_{z \sim \mathcal{D}_S} [\log (1 - D(G(z)))]
\end{aligned}
\end{equation}

\subsection{Semantic Discretizer}
\label{sec:Semantic Discretizer}
% \subsection{Semantic Discretizer}
In large-scale streaming training settings, in-batch negative sampling strategy (i.e. Eq.~\ref{in_batch_softmax}) is widely adopted  to train for  balancing computational efficiency \cite{hidasi2015session}. However, this approach inherently introduces a \textit{sampling bias}: popular items are more likely sampled as negatives, leading the model to over-penalize popular items. 
The mainstream industrial solution to mitigate this is $LogQ$ Correction \cite{yi2019sampling}, which adjusts the logits by subtracting $\log(P_i)$, where $P_i$ denotes the streaming frequency of item $i$.
Unfortunately, this approach falls short for active user queries; unlike templated inputs, manually typed queries are virtually unique. As a result, standard item-ID frequency statistics become meaningless (i.e., $P_i \approx \frac{1}{N}$), rendering traditional debiasing methods infeasible.

In Sec.~\ref{sec:Distribution Alignment}, we projected the active queries and starters into a unified latent space. Therefore, we could perform probability density sampling within this space to achieve debiasing. However, accurately estimating probability density in high-dimensional continuous representations is computationally intractable and unstable for streaming training.

To address this, we propose a \textbf{Semantic Discretizer} based on the Residual Quantized Variational Autoencoder (RQ-VAE).  The core idea is to bypass the difficulty of direct density estimation by discretizing the semantic space into clusters, allowing the cluster frequencies to serve as a proxy for probability density. Specifically, the RQ-VAE  quantizes the continuous text embedding $\mathbf{z}$ into a sequence of discrete code indices $c$ from a learned Codebook.  Besides, the streaming frequency $p(c)$ of each code ID  serves as probability density. We formulate  our \textit{Semantic-Debiased Loss} as:

\begin{equation}
\begin{split}
c_q= & \{c_1,c_2,c_3\}=RQVAE(z),\beta_q=1-D(G(z))\\
% P(c_q)=<p(c_1),p(c_2),p(c_3)>\\
\mathcal{L}_{debiased} &= - \beta_{q}\log \frac{\exp(f(u, q) - \log P(c_q))}{\sum_{q' \in \mathcal{B}} \exp(f(u, q') - \log P(c_{q'}))},
\end{split}
\end{equation}
where $z$ is the encoded query representation,  $c_q$ denotes the discrete semantic code assigned to query $q$. Moreover, $\beta_q$ represents the weight of current sample, which is the classification probability of the $D$. By incorporating the logit adjustment term $-\log P(c_q)$, we successfully mitigate the bias towards popular active queries and enhance the surfacing of long-tail intents.

\noindent\textbf{Multi-Task Learning.} Ultimately, to optimize our model, we jointly train the adversarial loss $\mathcal{L}_{\text{adv}}$ and the main task loss $\mathcal{L}_{\text{debiased}}$ within a multi-task learning framework. The final overall training objective can be formulated as:
\begin{equation}
    % 在这里替换成你的真实公式，例如：
    \mathcal{L}_{\text{total}} = \mathcal{L}_{\text{debiased}} + \lambda \mathcal{L}_{\text{adv}}.
\end{equation}

\section{Experiments}
% In this section, we evaluate the effectiveness of \modelname{} through both extensive offline experiments and online A/B testing.
\subsection{Experimental Setup}
\textbf{Dataset.} Given the lack of public datasets containing both active queries and starters, we construct a large-scale industrial dataset from our platform's production 90 Days' logs, which contain 7.2 billion samples.

\noindent \textbf{Evaluation Strategy.} To alleviate the feedback loops, in this work we introduce the user manually typed active queries to training. Crucially, to ensure a fair evaluation consistent with the actual deployment scenario, the test set is strictly restricted to starters from the candidate corpus $\mathcal{Q}^s$. In this work, we adopt the typical AUC as the  standard offline metric.

\noindent \textbf{Baselines:}
To ensure a fair comparison, all baseline models are based on a two-tower architecture and employ the same user tower architecture and feature set as \modelname{}. Our evaluation focuses on comparing different strategies for item modeling:
    % \item \textbf{ID-based Retrieval}: The standard industrial solution that assigns a unique learnable embedding ID to each query. 
\textbf{(1)Semantic Retrieval}: The Basic model described in Sec. 3.2. But it is trained exclusively on starter interaction logs , without incorporating active queries. \textbf{(2)ID-Based Retrieval}: The standard industrial solution that assigns a unique learnable embedding ID to each query. 
    \textbf{(2) Basic + Active}: A naive data augmentation baseline that directly incorporates active queries as positive samples.
     \textbf{(3) \modelname{} w/o Alignment}: An ablation variant where the adversarial distribution alignment module is removed.
   \textbf{(4) \modelname{} w/o Reweight}: An ablation variant that removes the reweight (i.e.$\beta_q$).

\begin{table}[htbp]
\centering
\setlength{\abovecaptionskip}{0pt} 
\setlength{\belowcaptionskip}{10pt}
\caption{Experiment results. The best results are highlighted in \textbf{bold}, and the second best are \underline{underlined}. }
\label{tab:offline_results_updated}
\small
\begin{tabular}{lcc}
\toprule
\textbf{Model} & \textbf{AUC (\%)} & \textbf{Rel. Imp. (\%)} \\ 
\midrule
ID-based Retrieval              & 59.76\% & -1.35\% \\
Semantic Retrieval (Basic)      & 60.58\% & 0.00\% \\
Basic + Active                  & 58.94\% & -2.71\% \\ 
\midrule
\modelname{} w/o Alignment      & \underline{60.71\%} & +0.21\% \\
\modelname{} w/o Reweight       & 58.00\% & -4.26\% \\
\textbf{\modelname{} (Ours)}    & \textbf{64.65\%} & \textbf{+6.72\%} \\ 
\bottomrule
\end{tabular}
\end{table}

\subsection{Offline Evaluation}
 The experimental results are shown in Table ~\ref{tab:offline_results_updated}. Our key observations are as follows:
 (1) \textbf{Limitations of Naive Augmentation:} The \textit{Basic + Active} and \textit{PA-Bridge w/o Alignment}  baselines fail to yield significant improvements over the basic semantic model.  This is attributed to the significant distribution bias between active queries and Starters. Simply mixing active queries into the training data without alignment fails to transfer knowledge effectively.
(2) \textbf{Superiority of Semantic Modeling:} Semantic-based methods significantly outperform the traditional ID-based baseline. We attribute this to the data sparsity issue. Moreover the ID-based method  tends to overfit historical feedback loops. In contrast, by leveraging rich textual content, semantic models demonstrate much stronger generalization capabilities.  
(3) \textbf{Efficacy of Proposed Components:} Comparing the ablation variants (\textit{w/o Alignment} and \textit{w/o Reweight}) with the full \modelname{}, we observe that the complete model achieves the best performance. This demonstrates that the Distribution Aligner and Semantic Discretizer are both indispensable, effectively addressing the twin challenges of cross-domain heterogeneity and the infeasibility of ID-based debiasing, respectively.

\begin{table}[htbp]
\small
\setlength{\abovecaptionskip}{0pt} 
    \setlength{\belowcaptionskip}{0pt}
\centering
\caption{Online A/B testing and data analysis results}
\label{tab:online_results}
\begin{tabular}{lc}
\toprule
\textbf{Metric} & \textbf{Relative Improvement ($\Delta$)} \\ 
\midrule
User Active Days        & $+0.04\%$ \\
Feature Penetration Rate & $+0.54\%$ \\ 
\midrule
\midrule
Unique Queries Exposed        & $+30.65\%$ \\
Unique Queries Clicked  & $+56.50\%$ \\ 
\bottomrule
\end{tabular}
\end{table}

\vspace{-3pt}
\subsection{Online A/B Testing}

To validate the effectiveness of \modelname{} in a real-world production environment, we conducted rigorous online A/B testing on our platform. We selected three core metrics to comprehensively evaluate the business value: (1) \textbf{User Active Days}, representing overall platform retention; and (2) \textbf{Feature Penetration Rate}, measuring the proportion of users who use conversation starter.
The results of the online A/B testing are illustrated in Table~\ref{tab:online_results}. Our method achieved a \textbf{+0.04\%} increase in User Active Days. It is important to note that for a mature application with such a massive user base, an improvement of this magnitude is considered highly significant. Furthermore, the proposed model drove  a \textbf{+0.54\%} uplift in  Penetration Rate, demonstrating its effectiveness in enhancing user engagement.

\subsection{Analysis of Breaking Feedback-loops}
\label{fig:anlaysis}
% \begin{figure}
%     \setlength{\abovecaptionskip}{0pt} 
%     \setlength{\belowcaptionskip}{0pt}
%     \centering
%     \includegraphics[width=0.8\linewidth]{experiment_analysis.png}
%     \caption{Starter click and exposure statistics }
%     \label{fig:placeholder}
% \end{figure}
In this work, we aim to utilize active user expressions to break historical feedback loops, thereby improving overall model performance. To further demonstrate the effectiveness of our PA-Bridge, we measured the diversity of our online results by tracking the number of unique queries exposed and clicked. As shown in Table~\ref{tab:online_results}, both metrics achieve significantly improvements. Notably, the unique impressed queries increase by \textbf{30.65\%}, demonstrating that incorporating active queries enables the system to expose diverse  queries and effectively \textbf{\textit{mitigate the feedback loop}}. More importantly, the unique clicked queries yield a remarkable \textbf{56.50\%} uplift.  This further demonstrates that incorporating active queries effectively enhances recommendation performance. 
%In reality, it is not that the candidate pool lacks starters aligned with user preferences; rather, users are harmed by historical feedback loops. These loops restrict them to selecting merely from a narrowly ranked list of system exposures, thereby preventing them from effectively expressing their true interests.
\section{Related Work}
Our research is closely related to the field of query recommendation. Traditionally, query suggestion has been extensively explored in standard search engines, encompassing tasks such as query autocompletion and query clarification. For instance, MPC~\cite{bar2011context} suggests top queries based on user input prefixes. To enhance semantic understanding, neural networks—particularly recurrent neural networks and attention-based models ~\cite{dehghani2017learning,mustar2020using,mustar2021study,sordoni2015hierarchical} have been widely adopted. Moreover, recent works like Onesug ~\cite{guo2025onesug} propose an end-to-end framework to replace traditional cascade architectures, while GQR~\cite{min2025prompting} utilizes LLMs to directly generate suggested queries.
Recently, as conversational search rapidly evolves, the focus has shifted towards proactive interactions, such as generating follow-up questions  \cite{li2025proactive, yin2025clicks}, clarifying ambiguous queries  \cite{deng2023prompting}, or steering topics  \cite{rebedea2024canttalkaboutthis}. However, most existing works focus on in-conversation stages, whereas we explore the before-conversation stage (i.e., conversation starter). Furthermore, while some recent methods rely on generative frameworks, in this work, we focus on the retrieval-based paradigm for high efficiency.

\section{Conclusion}
In this work, we focus on the emerging task of conversation starter recommendation within conversational search. To break the harmful phenomenon of historical feedback loops, we innovatively introduce users' active expressions into the recommendation process. 
% which effectively encapsulate their open-world query intents. 
Furthermore, to address the distribution shift and the non-ID-able problem, we design a novel framework named PA-Bridge, which adopt an adversarial learning framework and an RQ-VAE module to tackle these challenges. Online A/B testing demonstrates the effectiveness of our approach, yielding  0.54\% improvement in the module penetration rate. Now it has been deployed on our system.
\bibliographystyle{ACM-Reference-Format}
\balance
\bibliography{sample-base}

% \appendix
% \section{BIOGRAPHY} Yiqing Wu  and Haoming Li  is currently a Senior Engineer at ByteDance. \textbf{Yiqing Wu } received his Ph.D. degree from the Chinese Academy of Sciences. His main research interests include recommender systems, search, and graph.
% \textbf{Haoming Li} , received his Master's degree from the Chinese Academy of Sciences. His main research on recommender systems.

\end{document}